\def\be{\begin{equation}}
\def\te{\end{equation}}
\def\bea{\begin{eqnarray}}

\def\tea{\end{eqnarray}}

\def\d{\delta}

\def\G{\Gamma}

\newskip\humongous \humongous=0pt plus 1000pt minus 1000pt

\newif\ifdtup

\def\ha{{1\over 2}}

\documentstyle[12pt]{article}

\makeatletter                    
\@addtoreset{equation}{section}  
\makeatother                     

\begin{document}
\title{Noise and Fluctuations in Semiclassical Gravity}
\author{Esteban Calzetta\\
{\small IAFE and FCEN, Buenos Aires, Argentina}\\
B. L. Hu\\
{\small Department of Physics, University of Maryland, College Park,
MD 20742}}
\date{\today}
\maketitle
\centerline{(umdpp 93-216)}

\begin{abstract}

We continue our earlier investigation of the backreaction problem in
semiclassical gravity with the Schwinger-Keldysh
or closed-time-path (CTP) functional formalism using the language of the
decoherent history formulation of quantum mechanics.
Making use of its intimate relation with the
Feynman-Vernon influence functional (IF) method, we examine the
statistical mechanical meaning and show the interrelation of the
many quantum processes involved in the backreaction problem, such as
particle creation, decoherence and dissipation. We show how
noise and fluctuation arise naturally from the CTP formalism.
We derive an expression for the CTP effective action in terms of the
Bogolubov coefficients and show how noise is related to
the fluctuations in the number of particles created.
In so doing we have extended
the old framework of semiclassical gravity,
based on the mean field theory of Einstein equation
with a source given by the expectation value of the energy-momentum
tensor, to that based on a Langevin-type equation, where the dynamics of
fluctuations of spacetime is driven by the
quantum fluctuations of the matter field.
This generalized framework is useful for the investigation of quantum
processes in the early universe involving fluctuations, vacuum stability
and phase transtion phenomena and the non-equilibrium thermodynamics
of black holes. It is also essential to an understanding of the transition
from any quantum theory of gravity to classical general relativity.

\end{abstract}
\newpage
\section{Introduction and Summary}

The {\it central theme} of this paper is to show how the backreaction
problem in semiclassical gravity \cite{BirDav,scg} can be viewed in the light
of a quantum open system \cite{qos} and how
the concepts and techniques of non-equilibrium statistical
mechanics can be fruitfully applied to this theory for
the description of quantum statistical
processes in the early universe \cite{HuWaseda}.
This idea has been used recently to expound the dissipative nature of effective
quantum field theories \cite{HuPhysica,HuBanff}, some basic issues of
quantum cosmology \cite{HuTsukuba,HuErice,PazSin,HPS,HarLH} and
quantum mechanics \cite{HarDGR,ZurekPTP}.

The {\it primary aim} of this paper is to elevate the theory of
semiclassical gravity from the old level based on the semiclassical equation
with a source given by the vacuum expection value of the energy-momentum
tensor of quantum matter fields associated with particle creation
\cite{cpc} whose backreaction leads to dissipation
\cite{bkrncpc} in the dynamics of spacetime, to a new level
based on an Einstein-Langevin equation with a stochastic source given by
the fluctuations in the matter field, where the effects of
noise and fluctuations are also incorporated in the processes of decoherence
and dissipation.

The {\it main topics} of investigation reported in this paper
are noise and fluctuations in quantum fields associated with particle creation
in cosmological spacetimes; and dissipation in the dynamics of spacetime
due to the backreaction of these quantum processes.

The {\it specific findings} of this paper are:\
i) Explicitly showing the relation of particle creation with decoherence
through the noise kernel in the influence functional and the Bogolubov
coefficients in the theory of quantum fields in curved spacetime.
ii) Delineating the character of noise from the coupling of the
quantum field to the background spacetime.
iii) Deriving the noise terms (in addition to the average of
the energy-momentum tensor) in the semiclassical Einstein equation
as a stochastic source and relating the fluctuations of energy
density to the fluctuations in the number of particles created.

The {\it principal method} used here is that of the Schwinger-Keldysh
or closed-time-path (CTP) functional formalism \cite{ctp}.
This is the method we used (with a Bianchi Type-I universe as model)
\cite{CalHu87} in deriving a real and causal equation of motion for
the cosmological backreaction problem. From that we identified
a nonlocal kernel in the dissipative term and showed that the integrated
dissipative power in the dynamics of spacetime
is equal to the energy density of the total number of
particles created. This clearly established the dissipative nature of
quantum processes like particle creation \cite{CalHu89}. We now
describe the progression of ideas and  the evolution of the background
leading to the present work, which addresses the other part of this problem
(which actually
existed in our original results, but was not the focus of attention
in our earlier investigation) , i.e., noise and fluctuations.

Two earlier papers written by one of us outlined the usefulness of
adopting the quantum open system
point of view for understanding the dissipative nature
of quantum fields and semiclassical gravity \cite{HuPhysica} and some
basic issues of quantum cosmology \cite{HuTsukuba}.
Paper \cite{HuPhysica} noticed the missing role played by noise in
the equation of motion for the effective system, and advocated that a
Langevin-type equation should be used in place of the conventional
semiclassical Einstein equation. It was also predicted there
that for quantum fields under general conditions
a colored noise source should appear in the driving term. The other
two conjectures put forth in that paper, i.e., the existence of a fluctuation-
dissipation relation for non-equilibrium quantum systems which can be used
to understand backreaction problems in semiclassical gravity, and
the existence of dissipative behavior in effective field theories,
will be taken up in later investigations \cite{HuSin,CalHuFDR,HuBanff}.

Paper \cite{HuTsukuba} pointed out the interrelation
of quantum and statistical processes like
decoherence \cite{envdec,conhis,GelHar1,GelHar2,CalHuDCH,envhis,decQC},
correlation \cite{cor},
particle creation  (as amplification of vacuum fluctuations \cite{cpc}),
noise, fluctuation
\cite{Zhang,HPZ1,HPZ2,HuBelgium,HMLA,HM2,HabKan,BLM,LafMat,Mat},
dissipation \cite{HuErice,bkrncpc,CalHu87,CalHu89,Paz89,CalQC,HuSpain},
and their role in the evolution of the effective system
(which can be the classical limit of quantum mechanics \cite{GelHar2}
or the semiclassical spacetime dynamics from quantum cosmology
\cite{HarLH,PazSin}).
The pairwise relation of these processes have been explored since then
by many authors in various context. For example, that noise governs decoherence
was
seen in all the analysis of environment-induced decoherence \cite{envdec}.
This, together with the fluctuation-dissipation theorem which relates
noise to dissipation, implies that there is a limit to the degree of
decoherence and the accuracy of defining the classical trajectory
\cite{GelHar2}.
There is also a balance between decoherence and the build-up of correlation
between the canonical variables of a quantum system in reaching the
classical limit \cite{PazSin}.
The relation of particle creation and decoherence
is explored in \cite{CalMaz} as is also implicit in  \cite{PazSin}.
The relation of noise and fluctuations in particle number is studied
in \cite{HKM} for the quantum statistics of cosmological particle creation.

The main features of the quantum open system paradigm are well illustrated
by the quantum Brownian model. Using the Feynman-Vernon
\cite{FeyVer} formalism, and extending the work of
Caldeira and  Leggett  \cite{CalLeg83}  and  Grabert  et  al  \cite{Gra} to a
general environment, Hu,  Paz  and  Zhang  \cite{HPZ1,HPZ2}  looked  into the
nature of colored noise from the environment and the  nonlocal dissipation
they engender on  the dynamics of the system.  In this
formalism the effects of noise and  dissipation  can  be  extracted  from the
noise  and  dissipation  kernels  as the real  and  imaginary  parts  of  the
influence  functional,  their  interrelation  manisfesting  simply  as   the
fluctuation-dissipation theorem  \cite{fdr} obtained  as a categorical
functional relation.  If one views the quantum field as the environment
and spacetime as the  system  in the  quantum  open  system  paradigm,
then  the  statistical
mechanical meaning of the backreaction problem in semiclassical cosmology can
be understood more clearly \cite{HuPhysica}.  In particular, one can identify
noise  with  the  coarse-grained  quantum  fields, derive  the  semiclassical
Einstein  equation  as a Langevin equation, and understand  the  backreaction
process as the manifestation of a fluctuation-dissipation relation.

One gratifying by-product in this earlier process of search and discovery
is that the influence functional method \cite{FeyVer} used in the context of
non-equilibrium statistical mechanics is
largely equivalent to the Schwinger-Keldysh, or the closed-time-path (CTP)
method \cite{ctp} developed in quantum field theory.
This for us is particularly useful, because not only does one recover from
the IF the dissipation kernel in the equation of motion of the CTP, but one can
now clearly identify the meaning of the noise kernel already existent in the
CTP effective action and find the corresponding stochastic source in the
semiclassical equation of motion. We will indeed borrow the physical insight
provided by the IF formalism to  analyze the results obtained by the CTP
method.

The character and function of noise in some common quantum field processes
have been studied before in a different context. For example, \cite{HPZ1}
treated colored noise from a non-ohmic environment, \cite{HPZ2} dealt with
colored noise from nonlinear coupling, \cite{HM2} discussed particle creation
as the result of parametrically amplified quantum noise. The quantum
origin of noise and fluctuations basic to the gravitational-instability
theory of structure formation is discussed in \cite{HuBelgium}. The relation of
stochastic and thermal field is explained in \cite{qsf1} while that between
quantum noise and thermal radiance in accelerated observers and spacetimes
with horizons for the Unruh \cite{Unr} and Hawking \cite{Haw75,GibHaw}
effects is discussed in \cite{Sciama,Mottola,Ang,HMLA,HM2}.
We will discuss the backreaction problem in semiclassical gravity in
terms of the fluctuation-dissipation relation in later publications
\cite{HuSin,CalHuFDR}. Dissipation in
quantum cosmology arising from the neglected inhomogeneous modes in
a minisuperspace approximation leading to an effective Wheeler-DeWitt equation
was discussed in \cite{CalQC,SinHu}.
One could extract the noise corresponding to
the coarsed-grained modes of spacetime excitations and define a gravitational
entropy, as discussed in \cite{HuWaseda}. One could also deduce a
fluctuation-dissipation relation in quantum cosmology, exemplified by the
backreaction problem in Bianchi Type-1 universe  \cite{HuSin}.
Sharing the same goal as this paper but taking a different approach is
the work of \cite{HM3}, in which the colored noise associated with
quantum fields is identified by means of a cumulant expansion on the
influence functional and an Einstein-Langevin equation for
the backreaction problem was derived in semiclassical cosmology.
A recent paper of Kuo and Ford \cite{KuoFor} also addresses fluctuations
in semiclassical gravity. They work with the energy-momentum tensor
in the canonical formalism. Their approach and results should have
points of contact with ours (see Sec. 5 below).

The following is a brief description of the contents of
this paper.  In order to demonstrate
the stochasticity of semiclassical evolution
induced by quantum fluctuations, we shall analyze a
cosmological model in which a free, real scalar field is
coupled to the scale factor
of a Friedmann- Robertson- Walker (FRW) Universe.
One can think of this as the semiclassical limit of the corresponding
model in quantum cosmology, this transition having been studied by many
authors (the latest complete work is that of \cite{PazSin}, in which
are listed some of the earlier references).
Using the conceptual framework of the consistent or decoherent histories
approach
to quantum cosmology \cite{conhis,GelHar1}, we consider histories
where the matter field is
fully coarse grained. From this we obtain a closed, exact expression for
the decoherence functional between two such histories-- that is, between
two different specifications of the FRW conformal factor-- as a function
of time, this being the only remaining degree of freedom.
(This result was obtained also in \cite{PazSin,HM3}.) This expression
will allow us to show that decoherence is directly related to the differential
in particle creation between one and the other history.

{}From this we shall then discuss the dynamics of fluctuations in
the scale factor around its expectation value, as seen by an observer
who, by necessity or choice, is unaware of the presence of the scalar field
except for its overall effect on the dynamics of the system.
We shall show that this
dynamics is aptly described by a Langevin-type equation, where the usual
semiclassical corrections to the matter energy-momentum tensor are
supplemented by stochastic terms. Moreover, we shall deduce from the formalism
itself the noise auto-correlation function. We stress that this is the
correct way of treating fluctuations of quantum fields as noise.
Quantum fluctuations in the inflaton field viewed as seeds for galaxy
formation is a very attractive program
\cite{Sta,MFB}. But the existing practice is flawed
in at least two respects \cite{HuBelgium}: 1) The correct deduction of the
origin and nature of noise from quantum fluctuations; and 2)
The correct treatment of quantum to classical transition in
the long-wavelength perturbation modes via decoherence considerations.
We show how the quantum bath variables after average
effectively contribute a stochastic source with a correlation
function  determined by the nature of the bath and the coupling.
In the most general cases one expects colored and non-Gaussian noises
to appear.
The habitual way of simply reinterpreting the quantum scalar field
as a fluctuating classical Gaussian source with its mean square value
set equal to the corresponding quantum average value is incorrect except
at the coincidence limit.

As it can be seen from the above, the
hypothesis underlying our analysis is that
the nature of the semiclassical dynamics can only be appreciated in full by
combining concepts and techniques from quantum and statistical field theory.
For this reason, we shall begin in Sec. 2 with a brief summary
the closed time path effective action  \cite{ctp} and the
influence functional \cite{FeyVer} formalisms.
We shall show how the decoherence functional formulation provides a natural
framework for the application of these concepts to our problem.
In Section 3 we apply
these formalisms to the cosmological model described above,
arriving at an exact expression
for the decoherence functional in terms of the Bogolubov coefficients.
This expression makes obvious the connection between decoherence and
particle creation already mentioned.

In Section 4, we analyze the semiclassical dynamics experienced
by an observer confined to one of the decohering histories.
For definiteness, we shall focus on cosmic evolutions
which are close to a solution of the usual (deterministic)
semiclassical Einstein equations.
We shall show how the structure of the decoherence functional
implies the presence of noise in this dynamics, and determine
its statistical properties.
In this light we issue a warning that the usual procedure of treating
the spontaneous fluctuations in the field
as a classical Gaussian stochastic variable \cite{MFB} has only limited
validity.
In Section 4.3 we analyze the nonlocal
nature of noise and dissipation by examining a simple set of histories which
depart only slightly from the Minkowski space and discuss how the colored
nature of noise depends on the coupling of the field to spacetime.
In Section 5, we explain the physical origin of noise as fluctuations in
particle creation number. We show that the fluctuations
in the energy-momentum tensor calculated in the CTP formalism can also be
obtained from the fluctuations in the number of particles via simple
quantum field theory arguments.
In Section 6 we summarize our findings and discuss the implications of
our results.
A few technical details for the derivation of the main results in the text
are put in the Appendix.

\newpage
\section{Methods in Quantum and Statistical Field Theories}

As described in the Introduction, two methods have been used effectively
for the description of the backreaction problem:
the closed time path effective action (CTP, or Schwinger-Keldysh) formalism
\cite{ctp} for obtaining a causal and real equation of motion; and
the influence functional (IF, or Feynman-Vernon \cite{FeyVer})
method for treating a quantum open system, in identifying
the noise in the environment and the dissipation in the effective
equation of motion for the system.
We give here a brief description of these formalisms and their interconnection.
We also sketch the decoherent history formulation of quantum mechanics
as we will use this conceptual framework to apply  the IF and CTP formalisms
to the analysis of semiclassical gravity theory.

\subsection{The Influence Functional Approach to Nonequilibrium
Field Theory}

The IF approach \cite{FeyVer} is designed to deal with a situation in which the
system $S$ described, say, by the $x$ fields is interacting with
an environment $E$, described by the $q$ fields (in another common
statistical mechanical nomenclature these are also called the
relevant and irrelevant parts respectively).
The full quantum system is described by a density matrix
$\rho (x,q;x',q',t)$. If we are only interested in
the state of the system as influenced by the overall effect,
but not the precise state, of the environment,
then the reduced density matrix $\rho_r(x,x',t)=\int~dq~\rho (x,q;x',q,t)$
would provide the relevant information. (The subscript $r$ stands for
reduced.) It is propagated in time from $t_i$ by the propagator ${\cal J}_r$:

\be
\rho_r(x,x',t)
=\int\limits_{-\infty}^{+\infty}dx_i\int\limits_{-\infty}^{+\infty}dx'_i~
 {\cal J}_r(x,x',t~|~x_i,x'_i,t_i)~\rho_r(x_i,x'_i,t_i~)
\label{pathint}
\te

Assuming that the action
of the coupled system decomposes as $S=S_s[x]+S_e[q]+S_{int}[x,q]$,
and that the initial density matrix factorizes (i.e.,
takes the tensor product form),
$\rho (x,q;x',q',t_i)=\rho_s(x,x',t_i)\rho_e(q,q',t_i)$, the
propagator for the reduced density
matrix is given by

$$
{\cal J}_r(x,x',t|~x_i,x'_i,t_i)=
                     \int_{x_i}^{x_f} Dx~ \int_{x'_i}^{x'_f} Dx '~
e^{i(S_s[x]-S_s[x']+S_{IF}[x,x',t])},
$$
where $S_{IF}$ (called $\delta {\cal A}$ in \cite{HPZ1} ) is the
influence action related to the influence functional $\cal F$ defined by

\be
{\cal F}[x, x'] \equiv e^{iS_{IF}[x,x',t]}\equiv\int~dq_f~dq_i~ dq_i'~
  \int_{q_i}^{q_f}Dq~\int_{q'_i}^{q_f}Dq'~
e^{i(S_e[q]+S_{int}[x ,q]-S_e[q']-S_{int}[x ',q'])} \rho_e(q_i,q'_i,t_i).
\label{SIF}
\te
$S_{IF}$ is typically
complex; its real part ${\cal R}$, containing the dissipation kernel
$D$, contributes to the renormalization of $S_s$, and
yields the dissipative terms in the effective equations of motion.
The imaginary part ${\cal I}$,
containing the noise kernel $N$, provides the information about
the fluctuations induced on the system through its coupling to the environment.
Since
the connection between these kernels and their effect on the physical processes
of dissipation and fluctuation has been discussed at lenght elsewhere
(cfr. Ref.  \cite{HPZ1}), we shall  limit
ourselves here only to a schematic summary. \footnote{This
simplified schematic discussion is really just for the illustration of main
ideas, not for precision and completeness. The reader is referred to
\cite{HPZ1,HPZ2,HuBelgium,HMLA,HM2} for details on the discussion of the
process of decoherence in quantum to classical transition,
the origin and nature of quantum noise, the fluctuation-dissipation relation
and the explicit derivations of the master, Fokker-Planck and Langevin
equations
for the models of a Brownian particle in a general environment and interacting
quantum fields in  cosmological spacetimes.}

The main  features  of  the  influence  action
follow from  the  elementary  properties  $S_{IF}(x,x')=-S_{IF}(x',x)^*$  and
$S_{IF}(x,x)=0$, which can  be  deduced from its definition Eq.  (\ref{SIF}),
and derived in the final analysis  from  the  unitarity  of the underlying
quantum theory of the closed system.
If we decompose $S_{IF}$ in its real and imaginary parts,
 $S_{IF}={\cal R}+i{\cal I}$, then
${\cal R}(x,x')=-{\cal R}(x',x)$,
${\cal I}(x,x')={\cal I}(x',x)$, and ${\cal R}(x,x)={\cal I}(x,x)=0$.
Keeping only
quadratic terms, we may write

\be
S_{IF}(x,x')=\int~dt~dt'~\{ {1\over 2}(x-x')(t)D(t,t')(x+x')(t')
+{i\over 2}(x-x')(t)N(t,t')(x-x')(t')\}
\label{SIFquad}
\te
where $D$ and $N$ stand for the real dissipation and noise kernels respectively
($D\equiv 2\eta$, $N\equiv 2\nu$, in the notations of \cite{HPZ1}).
It is convenient to express $D$ as
$D(t,t')=-\partial_{t'}\gamma (t,t')$, and rewrite

\be
S_{IF}(x,x')=\int~dt~dt'~\{ {1\over 2}(x-x')(t)\gamma (t,t')(\dot x+\dot x')
(t')+{i\over 2}(x-x')(t)N(t,t')(x-x')(t')\}
\te
The physical meaning of  the $\gamma$ kernel may be elucidated by deriving
the mean field equation of motion for the mean value of the system variable
$\bar x$. It is

\be
{\partial S_s\over\partial\bar x(t)}+\int~dt'~\gamma (t,t')
{d\bar x(t')\over dt'}=0
\te
The term  containing $\gamma$
represents the  backreaction  of  the  environment  on  the
system.  It causes the dissipation  of energy from the system by an amount
(integrated over the whole history of the system)

\be
\Delta E=\int ~dt~dt'~\gamma (t,t')\dot {\bar x}(t)\dot {\bar x}(t').
\te
Thus we see that the even part of the kernel $\gamma$ is associated with
dissipation, while the odd part can be assimilated to a
nondissipative environment-induced change in the system dynamics.
In quantum field-theoretic applications, the odd part of $\gamma$ will
contain formally infinite terms which can  be  absorbed  in the classical
action for the system via standard renormalization procedures
\cite{ctp}. For  simplicity, we shall assume
that only the even part of  $\gamma$  is  left after renormalization has been
carried out.

In  general, the $\gamma$ and $N$ kernels are nonlocal;    however,  their
main  features    are  manifest  already  under  the  local  approximation
$\gamma\sim\gamma_0\delta (t-t')$, $N\sim N_0\delta (t-t')$.  The influence
action then takes the form

\be
S_{IF}(x,x')=\int~dt~\{ {1\over 2}(x-x')(t)\gamma_0(\dot x+\dot x')
(t)+{i\over 2}(x-x')(t)N_0(x-x')(t)\}
\te

Assuming an action functional of
the  simple  form  $S_s[x]\sim\int\{\ha    \dot    x^2-V(x)\}$,    it   is
straightforward  to derive the master  equation  for  the  reduced  density
matrix \cite{FeyVer,CalLeg83}

\be
i{\partial\rho_r\over\partial t}\sim
\{ [-\ha \partial_x^2+V(x)]-[-\ha \partial_{x'}^2+V(x')]
-i{\gamma_0\over 2}(x-x')[{\partial\over\partial x}-{\partial\over\partial x'}]
-i {N_0\over 2} (x-x')^2\}\rho_r
\label{master}
\te

The object `closest' (see \cite{cor}) to a classical distribution function
is the Wigner function \cite{Wigner}

\be
f_W(X,p)=\int~dy~e^{ipy}\rho_r(X+{y\over 2},X-{y\over 2})
\label{defwigner}
\te
where $X \equiv (1/2)(x + x'), y \equiv x - x'$.
The master equation (\ref{master}) implies (to lowest order in a
Kramers-Moyal expansion) the Fokker - Planck equation
\cite{Chandra}

\be
\{{\partial\over\partial t}+p{\partial\over\partial X}-V'{\partial\over
\partial p}\}f_W=(\gamma_0 {\partial\over\partial p} p+
{N_0\over 2}{\partial^2\over\partial p^2})f_W
\label{Fokker}
\te
(where $V'=dV/dx$).
{}From this equation one can see clearly the stochasticity in the semiclassical
dynamics. However, it is better to defer further discussion to subsection
{\bf  2.3}  below, until we  have  introduced  the  notion  of
the  decoherence functional. Suffice it to observe here that the Fokker -
Planck equation admits the equilibrium solution

\be
f^{eq}_W\sim e^{-(2\gamma_0 /N_0)[(p^2/2)+V(x)]}
\te
from which a fluctuation-dissipation theorem
$N_0=2\gamma_0 \langle p^2\rangle_{eq}$
can be derived. If the environment acts as a heat bath, then $
\langle p^2\rangle_{eq}\sim k_BT$, and this reduces to the
Einstein-Kubo formula for the dispersion coefficient.

\subsection{The Closed-Time-Path Functional Formalism in
Quantum Field Theory}

In the CTP approach, our goal is not to follow the dynamics of the
full density matrix, or even the system part, but only the expectation
values of the fields as they unfold in time. This evolution is
governed by a real and causal equation of motion, which is obtained
from the CTP effective action by a variational principle.

Let $\psi$ be the fields in the theory, and $\bar\psi$ their expectation
values for any given initial states. Consider pairs of histories
$(\psi,\psi')$ defined on all spacetime, with the property that $\psi(T^0)
=\psi'(T^0)$ for a given very large time $T^0$ (in practice, we shall
implicitly take the limit $T^0\to\infty$). Assume for simplicity
(more general choices are also possible \cite{CalHu88}) that the fields
were originally in their vacuum state $\vert 0IN\rangle$.
Then we can introduce external sources $J,J'$, and construct the CTP generating
functional

\be
Z[J, J']~=~e^{iW[J,J']}~
=\langle 0IN\vert \tilde T(e^{-i\int J'\psi})
T(e^{i\int J\psi})\vert 0IN\rangle
\label{Wcan}
\te
where  $T$  ($\tilde T$)  stands  for  time (anti-time) ordering.
Observe  that  the
generating  functional  $W$  is  totally  defined  once the IN  state  $\vert
0IN\rangle$ is chosen and that $W\equiv 0$ whenever $J=J'$.
Now introduce the path integral representation
\be
Z[J, J']~=~e^{iW[J,J']}~
=\int~D\psi~D\psi'~e^{i(S[\psi]-S^*[\psi']+J\psi-J'\psi')}
\te
The expectation values can be obtained as

\be
\bar\psi={\d W\over\d J}, ~~~
\bar\psi'=-{\d W\over\d J'}
\label{eqmotion}
\te
The physically relevant situation under consideration corresponds to
setting $J=J'=0$.

The  CTP effective action is just the Legendre transform of $W$

\be
\G_{CTP} [\bar\psi,\bar\psi']=W[J,J']-J\bar\psi+J'\bar\psi'
\te
where now the sources are thought of as functionals of the background fields
$\bar\psi,\bar\psi'$. In particular, the equations of motion are the inverses
of
Eqs. (\ref{eqmotion})

\be
{\d \G_{CTP}\over\d \bar\psi}=-J, ~~~
{\d \G_{CTP}\over\d \bar\psi'}=J'
\te
The physical situations correspond to solutions of the homogeneous equations
at $\bar\psi =\bar\psi'$. These equations are real and causal. Moreover,
$\G_{CTP} [\bar\psi,\bar\psi']
=-\G_{CTP}^*[\bar\psi',\bar\psi]$, and $\G_{CTP} [\bar\psi,\bar\psi]\equiv 0$.
As  the  generating  functional  itself, the CTP effective action is  totally
defined once the initial quantum state is given.

To apply this formalism to the situation above, we should substitute
the $\psi$ field by the pair $(x,q)$. When the physical situation requires
treating the $x$ and $q$ fields asymmetrically,
as is the case when, say, only the system field $x$ is relevant,
we do not couple the $q$ field to an external source. (In a perturbative
evaluation of the CTP generating functional, this means discarding
all graphs with $q$ fields on some external leg.)
Comparing the path integral expression for the generating functional
with the IF approach described earlier, we find

\be
e^{iW[J,J']}=\int~Dx~ Dx'~
{}~e^{i(S_s[x]-S_s[x']+Jx-J'x'+S_{IF}[x,x',+\infty ])}
\te
Conversely, we may describe the influence action as the CTP effective
action for the quantum $q$ fields interacting with external c-number
$x$ fields specialized to the expectation values of its arguments.

In the semiclassical approximation, one can
neglect Feynman graphs containing closed $x$ field loops, corresponding to
quantum effects of the $x$ fields.
Then the path integral and the Legendre transformation may be
computed explicitly, yielding

\be
\G_{CTP} [x,x']\approx S_s[x]-S_s[x']+S_{IF}[x,x',+\infty ]
\label{GSIF}
\te
This equation shows the connection between the CTP effective action and
the influence functional. From this we may derive the semiclassical
equations of motion for the expectation values of the $x$ field. We
see that the noise kernel does not contribute to these equations,
because, it being even under the exchange of $x$ and $x'$, its
variation vanishes at the coincidence point. However, as we shall argue below,
and is also clear from the master equation point of view \cite{HPZ1},
the noise kernel determines the dynamics governing the deviations
from the expectation value.

As a simple example of the foregoing, let us consider a model where the
system variable $x$ is coupled to an array of environment coordinates
$\{q_n\}$,    the    action  being  $S[x,q_n]=S_s[x]+\sum_n\{  S_e[q_n]+
\int~dt~\Xi
[q_n]x\}$  (models of this kind were considered by Schwinger \cite{ctp}
in his analysis of quantum brownian motion, and by many authors afterwards).

The CTP effective action takes the form
$\G_{CTP}[x, x']=S_s[x]-S_s[x']+\G[x, x']$,
$\G$ being related to $S_{IF}$ through Eq. (\ref{GSIF}).
Keeping  only  quadratic  terms in the CTP effective  action, we write

\be
e^{i\Gamma [x,x']}=e^{i\int~dt~dt'~\{G_{++}(t,t')x(t)x(t')+
G_{+-}(t,t')x(t)x'(t')+G_{-+}(t,t')x'(t)x(t')+G_{--}(t,t')x'(t)x'(t')\}}
\label{quadctp}
\te
On  the  other  hand,  under the semiclassical approximation for  the
system variable, we find

\be
e^{i\Gamma [x,x']}=
\langle 0IN\vert\prod_n\tilde T(e^{-i\int~dt~\Xi [q_n]x'})
T(e^{i\int~dt~\Xi [q_n]x})\vert 0IN\rangle
\label{semiclassctp}
\te
Taking the
variational derivatives of these equations with respect to $x$ and $x'$
at $x=x'=0$, we find

\be
G_{++}(t,t')=\sum_ni\langle  0IN\vert    T(\Xi    [q_n(t)]\Xi[q_n(t')])
\vert 0IN\rangle
\te

\be
G_{+-}(t,t')=\sum_n(-i)\langle  0IN\vert    (\Xi    [q_n(t')]\Xi[q_n(t)])
\vert 0IN\rangle
\te

\be
G_{-+}(t,t')=\sum_n(-i)\langle  0IN\vert    (\Xi    [q_n(t)]\Xi[q_n(t')])
\vert 0IN\rangle
\te

\be
G_{--}(t,t')=\sum_ni\langle  0IN\vert\tilde    T(\Xi    [q_n(t)]\Xi[q_n(t')])
\vert 0IN\rangle
\te
Introducing the kernels

\be
G(t,t')=\sum_ni\langle  0IN\vert    [\Xi    [q_n(t)],\Xi[q_n(t')]]
\vert 0IN\rangle
\te

\be
G_1(t,t')=\sum_n\langle  0IN\vert   \{(\Xi    [q_n(t)],\Xi[q_n(t')]\}
\vert 0IN\rangle ,
\te
where, as usual, square (curly) brackets denote (anti) commutators, we find

\be
\Gamma [x,x']=\int~dt~dt'~\{[x-x'](t)G(t,t')\theta (t-t')[x+x'](t')
+(i/2)[x-x'](t)G_1(t,t')[x-x'](t')\} ,
\te
($\theta$ being the step function)
which assumes the same pattern discussed above in the framework
of  the  influence  action  approach  (cfr. Eq. (\ref{SIFquad}),  after
identifying $D=2G$ and $N=G_1$).

\subsection{The Consistent Histories Approach to Quantum Mechanics}

Let us  now  relate these concepts and techniques in
statistical field theory to the more recent studies of the quantum to
classical transition
problem via the consistent histories formulation of quantum
mechanics \cite{conhis,GelHar1,GelHar2}.

In the consistent  or decoherent histories approach, the  complete description
of a coupled $x,q$ system is given in terms of fine-grained histories
$x(t),q(t)$. These histories are quantum in nature, i. e. it is possible
in principle to observe
interference effects between
different generic histories. A classical description is
acceptable only at the level of coarse-grained histories, and to the
extent that interference effects between these histories become
unobservable. Let us adopt the simple coarse-graining procedure
of leaving the $q$ field unspecified. Then each coarse-grained
history is labelled by a possible evolution of the $x$ field,
and the interference effects between histories are measured
by the decoherence functional (DF)

\be
{{\cal D}}[x,x']=
{}~e^{i(S_s[x]-S_s[x'])}
\int dq_i~dq'_i ~dq_f~\int Dq~~ Dq'~
e^{i(S_e[q]+S_{int}[x,q]-S_e[q']-S_{int}[x',q'])}
\rho_e(q_i,q'_i,t_i)
\te
which  is  the  fundamental  object  of  the theory. (For a more formal
definition see \cite{conhis,GelHar1}.)
The coarse-grained histories $x(t)$ can be described classically if
and only if the decoherence functional is approximately diagonal, that is,
${\cal D}[x,x'] \simeq 0$ whenever $x\not= x'$.  The conditions leading to this
in quantum mechanics is the focus of many current studies, to which we refer
the readers for the details. For quantum cosmology the issue is complicated
by the problem of time, and there even the definition of the decoherence
functional can be ambiguous \cite{HarLH}. In the problem of transition
from quantum cosmology  to semiclassical gravity, a WKB time is usually
assumed. In a work thematically related to this Paz and Sinha \cite{PazSin}
showed that an influence functional appears naturally from a reduced
density matrix by tracing out the matter fields. They discussed the decoherence
between WKB branches of the wave function and tried to relate it to the
notion of decoherence between spacetime histories. We assume in this work
that this essential step can be taken in some satisfactory way
and start our discussion at the semiclassical gravity level with the form of
the decoherence functional
\footnote{In coarse-graining away the environment
variable $q$ as in the simple Calderia-Leggett type models \cite{envdec},
there is no decoherence in the decoherent history sense \cite{conhis,GelHar1}
unless one makes a further coarse-graining of $x(t)$, such as specifying
the ranges of values of $x$ at different times. This is necessary
to ensure the consistency or decoherence condition which requires the
validity of the probability sum-rules for a set of histories. For the
condition for {\it a set of} histories to decohere is that the non-diagonal
elements of the decoherence functional vanishe {\it for all} pairs
of histories in the set. This extra coarse-graining on $x$ was explained
in \cite{GelHar2,envhis}. In so doing the simple form of the decoherence
functional (2.29) may become more complicated than necessary for the analysis
of
the semiclassical gravity domain. However, Gell-Mann and Hartle \cite{GelHar2}
had offered a partial solution to this problem, which we will assume for the
purpose of using the decoherence functional in the
semiclassical gravity form. We thank Juan Pablo Paz for calling our
attention to this point.}

\be
{\cal D}[x,x']=
e^{i(S_s[x]-S_s[x']+S_{IF}[x,x',\infty])}=
e^{i\G_{CTP} [x,x']}
\te
Notice that aleady at this formal level decoherence can occur only when
the noise kernel is nonzero, which signals the presence
of spontaneous fluctuations in the system.

We  now  arrive at the crucial point  of  our  analysis,  namely  the  proper
description  of  the  dynamics  of  a  single decohered  history
(that is, one particular decohered history chosen at random from
the heap of all possible consistent ones).
For an observer confined (by
necessity or by choice) to the level of coarse grained descriptions,
dynamical evolution must be described in terms of mutually exclusive
histories, all interference
effects having been suppressed below the accuracy of his observation devices.
For example, if he chooses to describe the evolution of the system in terms
of its Wigner Function $f_W$ (introduced in Section {\bf 2.1}), he will
now interpret it as an actual ensemble average, describing the
joint  evolution  of  the    bundle    of    coarse-grained   histories.
Correspondingly,
he will regard Eq. (\ref{Fokker}) as a classical Fokker-Planck
equation. Now the classical random process
described  by  Eq.   (\ref{Fokker})  is  not  deterministic;    rather,  it
describes the evolution of an ensemble of particles whose individual orbits
obey the Langevin-type equations

\be
\dot x=p~~~~~;~~~~\dot p=-V'-\gamma p+ \xi
\label{Langevin}
\te
where $\xi$ represents a noise term with autocorrelation
$\langle \xi (t)\xi (t')\rangle = N(t,t')$. (The ordinarily assummed
gaussian and white nature of the noise follows only from
a quadratic and local noise kernel, which describes rather special cases
in cosmological
situations, see \cite{HuBelgium,HMLA}). Thus, the observer confined to
a coarse-grained history will  conclude that
semiclassical evolution is stochastic.
Note that the statistical properties of this random evolution
are totally determined by
the  decoherence  functional  (or  equivalently, the  closed time path
effective action, or the influence functional); no {\it ad hoc} assumptions
on the behavior of quantum fluctuations are necessary.

As noticed by Feynman and collaborators (\cite{FeyVer}), there is a shortcut
to Eqs. (\ref{Langevin}): One can rewrite the part in the influence action
containing the noise kernel as

\be
e^{-\ha \int~dt~(x-x') N (x-x')}\equiv\int~D\xi~e^{i\int~dt~\xi(x-x')}
e^{-\ha \int~dt~\xi N^{-1}\xi}
\label{Foutrans}
\te
Therefore the action of the environment on the system may be described by
adding the external source term $-\int~x\xi$ to the system action $S_s$,
and averaging over external sources with the proper weight
\cite{Zhang,HPZ2,HM2}.
Variation of this effective action directly yields the Langevin equations
(\ref{Langevin}).
This is how noise can be understood
as a stochastic force from the environment acting on the system.

We are now ready to explore the consequences and implications of these
methods and ideas in the context of
semiclassical gravity. As a first observation, and in order to connect with
the more familiar language of quantum field theory in curved spacetime,
we show that decoherence and noise are
closely linked to particle creation, this being the main
dissipative mechanism in our problem.

\section{Decoherence Functional in terms of the Bogolubov Coefficients:
 Particle Creation and Decoherence}
We shall now
carry out an analysis of noise, fluctuations and dissipation
with the well-studied model of a Friedmann-Robertson-Walker (FRW) universe
filled with a quantum scalar field.

The metric for our model is

\be
ds^2=a^2(\eta )[-d\eta^2+\sum_{i=1}^{3}(dx^i)^2]
\te
where $\eta = \int dt / a$ is the conformal time.
(We assume spatial flatness only for definiteness, this plays
no role in the analysis below.) We shall use the  conventions of \cite{MTW}
throughout.

The scalar curvature for this model is

\be
R={2(n-1)\over a^2}\{{\ddot a\over a}+{(n-4)\over 2}({\dot a\over a})^2\}
\te
where a dot means a derivative with  respect to $\eta$, the conformal time, and
$n$ denotes
the spacetime dimension.
We are interested in the four dimensional case, of course, but for the
time being we may leave $n$ unspecified.

The Einstein-Hilbert action for general relativity is

\be
S_e={\rm constant}(-\int~d\eta~a^{n-4}\dot a^2)
\te
where the dimensional constant is $m_p^2(n-1)(n-2)L^{n-1}$,
this last factor being the ``volume'' of a surface of homogeneity, and $m_p$
the Planck's mass (in full consideration of renormalization, a
factor $\mu^{n-4}$, where $\mu$ is the renormalization scale,
should also be included).

Consider a real scalar free field with arbitrary mass $m$ and
coupling to curvature $\xi_n$. In terms of the canonical field variable
$\Phi$, the action is

\be
S_f=-{1 \over 2}\int~d^nx~\sqrt{-g}\{g^{\mu \nu}\partial_{\mu}\Phi
\partial_{\nu}\Phi +(m^2+\xi_n R)\Phi^2\}
\te
Specializing to our model, introducing the conformally-related field variable
$\Phi =a^{-1}\phi$, and discarding some total derivatives, we find

\be
S_f=\ha \int~d^nx~\{\dot\phi^2-(\nabla\phi )^2-M^2\phi^2\}
\te
where
\be
M^2=[m^2+(\xi_n -{(n-2) \over {4(n-1)}})R]a^2.
\te

{}From the discussions in Sec. 2.3 one can adopt the necessary procedures
linking semiclassical gravity with quantum cosmology
or follow the spirit of quantum field theory in curved spacetime
and begin the discussion of semiclassical gravity with (2.29).
Thus we assume that  the decoherence functional between
different histories $a_+(\eta ), a_-(\eta )$
of the conformal factor, after $\phi$ is totally coarse-grained away
takes the form

\be
{\cal D}[a_+,a_-]=\int~D\phi_+D\phi_-~e^{i([S_g[a_+]-S_g[a_-]
+S_f[a_+,\phi_+]-S_f[a_-,\phi_-])}
\te
Here in the gravitational action  $S_g=S_e+S_{tr}$ we have included
the trace anomaly-generating terms $S_{tr}$ arising from the
Jacobian of the $\Phi\to\phi$ transformation. As usual \cite{FHH}

\be
S_{tr}={L^{n-1}\over 2880\pi^2}\int~d\eta~\{-3({\ddot a\over a})^2
+({\dot a\over a})^4\}
\te

The histories are assumed to match at some point $\eta = \eta^o$ in the far
future, and the integration is over field histories such that
$\phi_+(\eta^o)=\phi_-(\eta^o)$. Further, we must choose the boundary
conditions
(and/or the measure) in the distant past to ensure convergence of the
path integral. For the purpose of this note, we shall adopt the simplest
procedure of assuming that for either evolution $a_+$ and $a_-$, $M^2$
vanishes in the distant past. Thus the boundary conditions can be fixed by
the same procedure as in a flat space time path integral, where again
we shall use the simplest criterion of tilting the path of integration
in the complex $\eta$ plane, in such a way that the $+$ branch acquires
a negative slope, and the $-$ branch a positive slope. If we think of
the integration path as a closed time loop, going from past to future on
the $+$ branch, and returning on the $-$ one, this means that the
imaginary part of $\eta$ is non increasing throughout \cite{Mills}.

To continue, let us decompose the field in plane waves (or
other spatial modes compatible with the symmetry of space).

\be
\phi (\vec x,t)=\int~{d^{n-1} \vec k\over (2\pi )^{n-1}}~e^{i\vec k \vec x}
\phi_{\vec k}(t)
\te
where  $k=\vert\vec k\vert$.
The amplitude of the $\vec k$-th mode obeys a wave equation of the type
\be
\ddot\phi_{\vec k}+\omega_k^2\phi_{\vec k} =0
\label{Kleingordon}
\te
where $\omega_k^2=k^2+M^2$. We shall omit the subindices $\vec k$ henceforth.

As is well known\cite{BirDav,cpc},
the quantization of the scalar field proceeds by further
decomposing each  Fourier  amplitude  in  its positive and negative frequency
parts, defined by  a suitable choice of time parameter.  This is accomplished
by developing the corresponding  mode  on a basis of solutions of the Klein -
Gordon equation Eq. (\ref{Kleingordon}), so normalized that the positive
frequency function has unit Klein  -  Gordon  norm,  the  negative  frequency
function has norm $-1$, and they  are  mutually  orthogonal  in  the  Klein -
Gordon inner product.  Such a basis of solutions constitute a particle model.
Properly  normalized  particle  models  are  related to  each  other  through
Bogolubov  transformations.    Let  us  observe that, each  function  of  the
particle  model  being a solution of Eq.  (\ref{Kleingordon}),  the  particle
model  may be defined by simply giving the corresponding Cauchy  data  on  an
arbitrary Cauchy surface. Further identification of the coefficient of the
positive frequency function in the development of the field, as a destruction
operator, allows for the second quantization of the theory. The particle
model is also associated to a vacuum state, which is  the  single common null
eigenvector of the destruction operators, and to a Fock basis, built from the
vacuum through the action of the creation operators

It  is  also well known that in a generic dynamic space  time,  there  is  no
single  particle  model  which  can  be identified outright with the physical
concept of  ``particle'';    however,  oftentimes  it is possible to employ a
variety of criteria  (such as minimization of the particle number as detected
by a free falling  particle  detector, Hamiltonian diagonalization, conformal
invariance, analytical properties in the euclidean section of the space time,
if any, etc) to single out a preferred particle model in the distant past (or
``IN'' particle model), and another in  the  far  future, or ``OUT'' particle
model.  In general, these models are not  equivalent,  the
vacuum of one model being a multiparticle state in the other.

In our problem, the choice of boundary conditions for the path integral above
amounts  to  a definite choice of the IN particle model, and the  IN  quantum
state, in each branch of the closed time path.  Indeed, because $M^2\to 0$ in
the distant past  on  either  branch, the field becomes conformally invariant
there, so that quantization can be carried out as in Minkowsky
space time.  Now our  procedure of deforming the time path into
the complex plane
would pick up the Minkowsky vacuum;    so  in  a  generic  spacetime,  we are
defining the initial state to be the conformal vacuum, and the IN particle
model
to be the conformal one.  As shown in the previous section, the choice
of initial state defines the CTP effective action.
Making the provisos discussed there we may write

\be
{\cal D}[a_+,a_-]=~e^{i\{ S_g[a_+]-S_g[a_-]+\Gamma [a_+,a_-]\}}
\label{decoeffac}
\te
where  $\Gamma$  is  the influence or effective  action  for  the  scalar
field,
evaluated at vanishing field background, and the
conformal  factor being treated as an external field.

Since the CTP effective action is independent of the OUT quantum states, we
have more freedom in choosing an OUT particle model. It is convenient to
choose a common OUT particle model for both
evolutions (that is, the Cauchy data on the matching surface $\eta=\eta^o$
are the same although the actual basis functions will be different).
The positive-frequency time dependent amplitude functions
$f_{\pm}$ for the conformal model in each branch are
related to those $F$ of the OUT model by
$f_{\pm}=\alpha_{\pm}F+\beta_{\pm}F^*$ at $\eta=\eta^o$,
where $\alpha_{\pm},\beta_{\pm}$  are  the  Bogolubov  coefficients  in  each
branch,
obeying the normalization condition $|\alpha_{\pm}|^2-|\beta_{\pm}|^2=1$.
The CTP effective action in Eq. (\ref{decoeffac}) is found to be:

\be
\Gamma =({i\over 2})\ln [\alpha_-\alpha_+^*-\beta_-\beta_+^*]
\label{Gammabogo}
\te
We give two independent proofs of this formula in the Appendix (Sec. 1 and 2).
We also show that it leads to real and causal corrections to
Einstein's equations in Appendix Sec. 3. This expression is exact.
(A similar expression can be obtained
from the influence functional for cosmological models \cite{HM3}).

The lesson for us is that there can be decoherence
($Im \G > 0$) if and only if there is particle creation
in different amounts in each evolution. (This is also implicit in
\cite{PazSin}.) Indeed, we can always
choose the OUT model so that $\alpha_+=1$, $\beta_+=0$, yielding
$\Gamma =(i/2)\ln \alpha_-$. The condition for decoherence in this case is
then $|\alpha_-|> 1$. But since $|\alpha_-|^2=1+|\beta_-|^2$,
this can only happen if there is particle creation between these two
particle models.

For this simple model this result suggests that the physical
mechanism underlying decoherence in the decoherent history scheme of Gell-Mann
and Hartle \cite{conhis,GelHar1} is the same as in the
environment-induced scheme \cite{envdec} based on a reduced
density matrix obtained by projecting 
\cite{projop} from the full density matrix and
tracing over the environmental degrees of freedom.
(For the connection between these two schemes see \cite{envhis}). If
the system and environment are correlated (i.e., that the full density
matrix cannot be decomposed into a tensor product of system and
enviromment states), this tracing procedure will leave the system in a
mixed state.

The correlations between the system and the environment
may be present in the initial conditions,
or they may arise dynamically. Since in our initial
condition the system ($a$) and the environment ($\phi$) are uncorrelated,
decoherence occurs only when correlations are generated in the dynamics.
For free fields, as in this model, correlations between the scale factor and
the fields are generated through particle creation.
(For example, consider a combined tensor product quantum state where the
field is in its vacuum state for some value of the scale factor. Although
the field state would react to adiabatic
changes in $a$, the combined state will remain a tensor product unless
particle creation occurs.) The problems of correlations engendered
by particle creation and interaction and their role in entropy generation
have been considered in \cite{HuPav,HuKan,HKM}.

One may observe that since $\Gamma$ becomes identically zero
when its arguments coincide, one seems to get the same
probabilities for all coarse-grained histories.
In actuality this only means that further coarse-graining may be
necessary to obtain a set of histories compatible with
the actual description of our Universe.

\section{Equation of Motion, Noise and Fluctuation}

\subsection{Equation of Motion}

Recall from Sec. 2
that the expectation value of the conformal factor obeys
the equation
\be
{\d S_g\over\d a}+{\d\Gamma [a_+,a_-]\over \d a_+} \vert_{a_+=a_-=a}=0
\te
Being causal and nonlocal,
this equation cannot be derived from an action functional.
Let us now consider the dynamics of small fluctuations $\delta a_{\pm}$
around a solution $a$ of the semiclassical equations above.
To do this, we shall
start by computing the CTP effective action for the field.

As in the previous section, we shall choose as particle model that which
reduces to the conformal model in the distant past. Since the unperturbed
evolution is the  same  on  either  branch  of  the  closed  time  path, this
condition defines a single  unperturbed  IN  particle model.  Projecting this
model
to the far future, we obtain also an OUT particle model.  We shall adopt this
choice, which reduces the unperturbed Bogolubov coefficients to 1 and 0.
Since we
just want the effective action up to quadratic order in the perturbations,
with this choice
we only need the perturbed $\beta_{\pm}$ coefficients to linear order,
and the perturbed $\alpha_{\pm}$ ones to second order.

Let $f$ be the positive frequency function of the unperturbed
particle model defined above, and let $f_{\pm}$ be the
positive frequency functions of the perturbed conformal particle models.
Then $f_{\pm}$ has an expansion
$f_{\pm}=f+f^I_{\pm}+f^{II}_{\pm}+...$ in powers of the perturbation (denoted
here by the super Roman numerals). By construction, the Cauchy data for
$f_{\pm}$ are the same as for $f$, therefore
the correction terms must vanish in the distant past.

Introduce the notation
\be
\Delta\omega^2_{\pm}=\int~d\eta '{\delta\omega^2\over\delta a(\eta ')}
\delta a_{\pm}(\eta ')
\te
for the correction to $\omega^2$ due to the perturbation. The identity
\be
{\delta\Delta\omega^2\over\delta\delta a(\eta )}=
{\delta\omega^2\over\delta a(\eta )}
\te
follows from this definition. In terms of $\Delta\omega^2$, we find
\be
f^I_{\pm}(\eta )=-\int^{\eta}_{-\infty}~d\eta '~G(\eta ,\eta ')
\Delta\omega^2_{\pm}(\eta ')f(\eta ')            
\te
where $G$ is the retarded propagator
\be
G(\eta ,\eta ')=i[
f^*(\eta ')f(\eta )-f(\eta ')f^*(\eta )]\theta (\eta -\eta ')
\te
which is of course independent of the actual choice of particle model.
Using the explicit expression for $G$, we obtain

\be
f^I_{\pm}=[1-i\int^{\eta}_{-\infty}~d\eta '~|f(\eta ')|^2
\Delta\omega^2_{\pm}(\eta ')]f(\eta )+
[i\int^{\eta}_{-\infty}~d\eta '~f(\eta ')^2
\Delta\omega^2_{\pm}(\eta ')]f^*(\eta )
\te
{}From this we can read  off  the  $\beta_{\pm}$  coefficients  to  the desired
accuracy

\be
\beta_{\pm}=i\int^{+\infty}_{-\infty}~d\eta ~f(\eta )^2
\Delta\omega^2_{\pm}(\eta )+{\rm O}[(\Delta\omega_{\pm}^2)^2].
\te
Iterating this procedure, we
get the $\alpha$ coefficients \cite{CalCas}

\begin{eqnarray}
\alpha_{\pm}=\{1+&\int^{+\infty}_{-\infty}~d\eta ~f^*(\eta )^2
\Delta\omega^2_{\pm}(\eta )\int^{\eta}_{-\infty}~d\eta '~f(\eta ')^2
\Delta\omega^2_{\pm}(\eta ')+{\rm O}[(\Delta\omega_{\pm}^2)^3]\}\cr
&{\rm exp}
[i\int^{+\infty}_{-\infty}~d\eta ~|f(\eta )|^2
\Delta\omega^2_{\pm}(\eta )]
\end{eqnarray}

Observe that indeed the normalization condition is satisfied.
Inserting these expressions back in the formula for the effective action,
we find a term of first order in the perturbation, which is cancelled by
the  variation  of  the classical action  (since  we  assume  the  background
evolution  is  a  solution  of  the semiclassical  equations  of  motion  for
the expectation values), and a quadratic term

\begin{eqnarray}
\Gamma [\delta a_+,\delta a_-]=(i/2)\{&
\int^{+\infty}_{-\infty}~d\eta ~f(\eta )^2
\Delta\omega^2_{+}(\eta )\int^{\eta}_{-\infty}~d\eta '~f^*(\eta ')^2
\Delta\omega^2_{+}(\eta ')\cr
+&\int^{+\infty}_{-\infty}~d\eta ~f^*(\eta )^2
\Delta\omega^2_{-}(\eta )\int^{\eta}_{-\infty}~d\eta '~f(\eta ')^2
\Delta\omega^2_{-}(\eta ')\cr
-&\int^{+\infty}_{-\infty}~d\eta ~f(\eta )^2
\Delta\omega^2_{-}(\eta )\int^{+\infty}_{-\infty}~d\eta '~f^*(\eta ')^2
\Delta\omega^2_{+}(\eta ')\}
\end{eqnarray}
leading to the equations of motion for the expectation value of the
perturbation

\be
\int~d\eta '~{\d^2S_g\over\d a(\eta )
\d a(\eta ')}\delta a(\eta ')+
\int~d\eta  '~{
\delta\omega^2(\eta ')\over\delta a(\eta )}
\int~d\eta ''~D(\eta ',\eta '')
\Delta\omega^2(\eta '')=0                     
\te
where
\be
D(\eta ',\eta '')={i\over 2}
[f(\eta ')^2f^*(\eta '')^2-f^*(\eta ')^2f(\eta '')^2]
\theta (\eta '-\eta '')
\te
As expected, this equation is real, causal and non local \cite{CalHu87}.
The boundary
conditions are that $\delta a$ must vanish in the distant past.
However, this is the equation only for the expectation value of the
perturbation, and since we are perturbing
around a solution of the semiclassical equations of motion, the only solution
with those boundary conditions is the trivial one.
What we want
to describe is the effective dynamics of
the conformal factors alone, which, as we discussed in detail in Section 2,
is stochastic in nature and described by a Langevin equation.
This equation, in turn, is best derived following Feynman's procedure
\cite{FeyVer}.
To implement Feynman's method, it is convenient to introduce the
symbols $\{ X\}\equiv (X_++X_-)$ and $[X]\equiv  (X_+-X_-)$.
In this notation, the effective action reads

\begin{eqnarray}
\Gamma =&\ha\int~d\eta~d\eta '~
[\Delta\omega^2(\eta )]
D(\eta ,\eta ')\{\Delta\omega^2(\eta ')\}\cr
&+{i\over 2}\int~d\eta~d\eta '~
[\Delta\omega^2(\eta )]
N(\eta ,\eta ')[\Delta\omega^2(\eta ')]
\end{eqnarray}
where

\be
N(\eta  ,\eta  ')={1\over  4}\{  f(\eta  )^2f^*(\eta  ')^2+f^*(\eta )^2f(\eta
')^2\}
\te
We can see a sort of ``division of labor'' here: the first term,
the dissipation kernel, determines
the equation of motion, but does not contribute to decoherence, while
the second term, the noise kernel, does not affect the equations of motion,
but is responsible for decoherence.

It may be argued that, if one summed over all modes, one could get
exact decoherence, since the decoherent terms could diverge
\cite{PazSin}. This effect is, however, generally believed to be unphysical.
Indeed, more physically relevant coarse graining strategies seem to avoid this
pitfall \cite{CalHuDCH}.

The failure of our observer to reduce the first term to a difference
between functionals of each history separately was expected, since we knew
that no such functional could lead to the proper equations of motion.
Physically, it is the dissipative nature of semiclassical evolution
which precludes its formulation in terms of an action principle. We shall
see an example of this below. However, an observer in the coarse-grained
history could still think of this first term as arising from both a classical
action and a dissipative function (see Ref. \cite{Landau}, entry 121).

\subsection{Noise}

To understand the meaning of the second term better, recall that the
decoherence functional has the form \cite{FeyVer}

\be
{\cal D}[\delta a_+,\delta a_-]=e^{
{i\over 2}\int~d\eta d\eta '[\Delta\omega^2](\eta )
D(\eta ,\eta ')\{\Delta\omega^2\}(\eta ')}
e^{-\ha \int~d\eta d\eta '[\Delta\omega^2](\eta )
N(\eta ,\eta ')[\Delta\omega^2](\eta ')}
\te
Performing the functional Fourier transform

\be
e^{-\ha \int~d\eta d\eta '[\Delta\omega^2](\eta )
N(\eta ,\eta ')[\Delta\omega^2](\eta ')}=\int~D\xi~{\cal P}[\xi]~
e^{-i\int~d\eta [\Delta\omega^2](\eta )\xi(\eta )}
\label{Fourier}
\te
the decoherence functional may be understood as the result of averaging
the functional

\be
e^{{i\over 2}\int~d\eta d\eta '[\Delta\omega^2](\eta )
D(\eta ,\eta ')\{\Delta\omega^2\}(\eta ')}
e^{-i\int~d\eta [\Delta\omega^2](\eta )
\xi(\eta )}
\te
over all possible values of a stochastic external source $\xi$,
with probability distribution ${\cal P}[\xi]$. To this order of expansion $\xi$
is a Gaussian variable
which produces a stochastic source on the right hand side of the equation
for $\delta a$, namely,
\begin{eqnarray}
\int~d\eta '~{\d^2S_g\over\d a(\eta ) \d a(\eta ')}\delta a(\eta ')
& +\int~d\eta '~{\delta \omega^2(\eta ')\over\delta a(\eta )}
\int~d\eta ''~D(\eta ',\eta '')\Delta\omega^2(\eta '')=\cr
& \int~d\eta '~{\delta \omega^2(\eta ')\over\delta a(\eta )}  \xi (\eta ')
\label{earlierresults}          
\end{eqnarray}

Due to the nonlocality of the noise kernel, the noise is generally
nonwhite; it is also generally non-Gaussian \cite{HuBelgium,HM3,CalHuNoise}.
Indeed, its Gaussian nature in our example
is merely a result of our having stopped at quadratic order
in the expansion of the effective action. The important thing to notice
here is that the formalism itself saves one the trouble (or embarrasment)
of making {\it ad hoc} and oftentimes inconsistent guesses about the
nature of the noise. For linear perturbations, the noise is Gaussian,
with auto-correlation

\be
C(\eta ,\eta ')=\int ~d\eta ''\int ~d\eta '''
{\delta \omega^2(\eta '') \over\delta a(\eta )}N(\eta '',\eta ''')
{\delta \omega^2(\eta ''')\over\delta a(\eta ')}
\label{sigma}
\te
$C$ is, of course, the expectation value of the product of the noise
at times $\eta$ and $\eta '$. Eq. (\ref{sigma}) follows from taking two
derivatives of Eq. (\ref{Fourier})
with respect to $\Delta\omega^2$, then setting
this to zero.

It is a remarkable fact that, for histories,
the more classical they become, the noisier they are. This point was
emphasized in \cite{GelHar2,PazSin}.
Mathematically this follows from decoherence and noise being determined
by the same kernel $N$.
Physically, in our context, it follows from the fact that noise and
decoherence are both
related to particle creation and backreaction.
Indeed, noise is just the difference between the stochastic process of
particles as they are actually created, and the
smoothed-out average effect represented by the expectation value.

\subsection{Fluctuations}
The assumption that quantum fluctuations in the fundamental fields can somehow
transmute into classical stochastic fluctuations is central to the stochastic
inflation program \cite{Sta}
and underlies most theories of galaxy formation via the perturbation of
quantum fields \cite{MFB}.  Though widely accepted and applied,
the crucial point in this program, this transmutation or transgression,
has never been satisfactorily proven. (For a critique of this view,
see \cite{HuBelgium}). The usual prescription
is to consider $ \phi$ as a classical
stochastic Gaussian variable, with an auto-correlation chosen to match
the quantum two point functions.
Let us see what can go wrong with  this {\it ad hoc} assumption.

It is helpful to again look at this problem from a slightly different angle.
As is well-known,
the single equation for the conformal factor we have
derived here is equivalent to the trace of the full Einstein
equations \cite{Par79}. More concretely, from the definition

\be
T_{\mu\nu}={2\over\sqrt -g}{\d S_f\over\d g^{\mu\nu}}
\te
one gets

\be
{\d S_f\over\d a}=-a^3T^{\mu}_{\mu}
\te
Therefore, as a Heisenberg operator, the trace of the energy momentum
tensor (not counting the trace anomaly terms already included in $S_g$)
is given by

\be
T^{\mu}_{\mu}(\eta )={1\over 2a^3}\int~d\eta '~{\delta \omega^2(\eta ')
\over {\delta a(\eta)}} \phi^2(\eta ')   
\te
Comparing these formulae to Eq. (\ref{earlierresults} )
we see that the stochastic source $\xi$ corresponds to the random
fluctuations $\ha ( \phi^2-\langle \phi^2\rangle_q)$, where
$\langle\rangle_q$ denotes the expectation value of an observable
with respect to the IN vacuum, computed from the usual rules in quantum
field theory. In our approach, the expectation value $\langle\phi^2\rangle_q$
is automatically included in the nonstochastic part of the
effective action.

Let $\langle\rangle_c$ denote the classical ensemble average over the different
values of the source. Since $\langle \xi\rangle_c=0$,
we find $\langle\phi^2\rangle_c=\langle\phi^2\rangle_q$. Moreover,
from $\langle \xi(\eta)\xi(\eta ')\rangle_c=N(\eta ,\eta')$, we find
\begin{eqnarray}
\langle \phi^2(\eta )\phi^2(\eta ')\rangle_c
-\langle \phi^2(\eta )\rangle_c\langle\phi^2(\eta ')\rangle_c
&=4N(\eta ,\eta ')\\
&=\{f(\eta )^2f^*(\eta ')^2+f^*(\eta )^2f(\eta ')^2\}
\end{eqnarray}
If we compare this result to the corresponding quantum average, we find
\be
\langle \phi^2(\eta )\phi^2(\eta ')\rangle_c
={1\over 2}\langle \{\phi^2(\eta )\phi^2(\eta ')
+\phi^2(\eta ')\phi^2(\eta )\}\rangle_q
\label{qtoclass}
\te
The need to symmetrize the quantum average could be expected, since
there is no analog of non-commuting variables in classical stochastic dynamics.

Before discussing further the implications of this equation,
let us try to recover this result by simply viewing the field $\phi$ as a
classical stochastic field, as is done in almost
all discussions on this subject (for a review e.g. \cite{MFB}).
As the auto-correlation $\langle \phi(\eta )
\phi(\eta ')\rangle$
should be real and even in its arguments, the
only choice is to identify it with the Hadamard function, yielding

\be
\langle \phi(\eta )  \phi(\eta ')\rangle_{c'} =\ha G_1(\eta ,\eta ')
               =\ha \{f(\eta )f^*(\eta ')+f^*(\eta )f(\eta ')\}
\te
(we use the subindex $c'$ to distinguish these
averages from those discussed above).
But then the Gaussian character of this variable implies

\be
\langle \phi^2(\eta ) \phi^2(\eta ')\rangle_{c'} -
\langle \phi^2(\eta )\rangle_{c'}\langle \phi^2(\eta ')\rangle_{c'}
=\ha G_1(\eta ,\eta ')^2
\label{Hadamard}
\te
which fails to reproduce the quantum average, even after symmetrization
[cf. Eq. (\ref{qtoclass})]. On this count it can be seen that
the conventional view on quantum fluctuations is flawed. (It also misses
out the full complexity of the issue of quantum to classical transtion, see
\cite{HuBelgium}).
Fortunately Eq. (\ref{Hadamard}) can yield the correct result
in the coincidence limit $\eta '=\eta$.
For this reason, the usual scenarios for
the generation of primordial fluctuations in inflationary cosmology
can remain valid if the proper form of the noise correlator is used.

The equality between the specific kinds of classical and quantum averages
defined in Eq. (\ref{qtoclass})
warrants that several familiar results from  quantum field theory in curved
spacetime will also be valid in the semiclassical approximation. For
example, the mean square value of the spontaneous fluctuations of a
massless, minimally coupled scalar
field in de Sitter spacetime will grow linearly with cosmological time
\cite{VilenkinFord}.    This  result  is  consistent with  the  view
that these fluctuations can be represented as white noise \cite{HMLA,HM2},
associated with the thermal radiation at the Hawking temperature
of the De Sitter universe \cite{Haw75,GibHaw}.
We shall demonstrate
this equivalence between quantum and semiclassical results for a more complex
example in Sec. 5.

On consideration of self-interacting quantum field theories, which  would not
be one -loop exact
(as is the case for the free field theory we are discussing here), the
quantum to classical correspondence would not necessarily hold beyond one
loop.    In these conditions, however, it is  possible  to  improve
systematically the accuracy of the semiclassical approximation by a
suitable choice of coarse graining procedure, as in the ``correlation
histories'' approach. We have discussed these issues elsewhere \cite
{CalHuDCH}.

\subsection{Nonlocal Kernels and Colored Noise}

In the above we have derived an expression for the dissipation $D$ and
noise kernel $N$ in terms of the positive frequency components
of the amplitude functions of the conformal IN
particle model of a given consistent cosmology. To examine
the structure of these kernels, we shall now spacialize to a
particularly simple but illustrative model, by choosing the background-
consistent evolution to be just the Minkowski spacetime. That is, we are
interested in the physics of small departures from the special case of
Robertson-Walker conformal factor $a^2=1$ under the influence of, say,
a real,
massive, free scalar field.

In this simple case, the unperturbed positive frequency modes
are just

\be
f_k(\eta )={1\over\sqrt{2\omega_k}}e^{-i\omega_k\eta}
\te
where the natural frequency is $\omega_k^2=k^2+m^2$,
(because of the spherical symmetry the modes can be labeled by
$k \equiv |\vec k|$).
The sum over all modes, with the dimensionless measure $Vd^3 \vec k/(2\pi )^3$
can be conveniently expressed in terms of
the positive frequency Wightman function

\be
G_+((\eta ,\vec x),(\eta ',\vec x'))=\int{d^3\vec k\over (2\pi )^3}f_{\vec k}
(\eta ) f^*_{\vec k}(\eta ')e^{i\vec k(\vec x-\vec x')}
\te
We find
\be
\int{Vd^3\vec k\over (2\pi )^3}f^2_{\vec k}(\eta ) f^{2*}_{\vec k}(\eta ')
=\int d^3\vec x d^3\vec x'~G_+^2((\eta ,\vec x),(\eta ',\vec x'))
\te
which, by virtue of translation invariance, becomes

\be
V\int d^3\vec x~G_+^2((\eta ,0),(\eta ',\vec x))
\te
The positive frequency Wightman function also admits the representation

\be
G_+(x^{\mu},{x'}^{\nu})=\int~{d^4p\over (2\pi )^4}e^{ip_{\mu}(x^\mu-x'^\mu)}
                                         2\pi\delta (p^2+m^2)\theta (p_0)
\te
from which we get the well-known relation,

\be
\int{Vd^3\vec k\over (2\pi )^3}f^2_{\vec k}(\eta )
f^{2*}_{\vec k}(\eta ')={V\over (4\pi )^2}\int~d\omega~e^{-i\omega
(\eta -\eta ')}
\sqrt{1-{4m^2\over\omega^2}}\theta (\omega -2m)
\te

In the massless case, the last integration is trivial, yielding

\be
{V\over (4\pi )^2}\{-iPV[{1\over\eta -\eta '}]+\pi\delta (\eta -\eta ')\}.
\te
Since the noise kernel is just one half of the real part of this expression,
we get immediately

\be
N(\eta ,\eta ')={V\over 32\pi}\delta (\eta -\eta ').
\te
Thus we see that a massless free field is associated with a purely white noise
(which makes physical sense, since there is no dimensionful scale to define a
memory time). This result is relevant more generally to the study of
near-conformal fields on arbitrary background Robertson-Walker spacetimes.
As long as the departure from conformal invariance is small, we can use
the conformal modes in the formal expressions from the previous subsections
which are equivalent to those of a Minkowski massless field.

For $m\not= 0$, the integral is not easily done, but we can reason
as follows (see Ref. \cite{Spino}): When the lapse
$\eta -\eta '$ is small, the integral will be dominated by
high frequencies. But in this regime, the mass is unimportant, so
we still get the delta function singularity. The mass begins to play a
role for finite time separation. In particular, for very large time
separations $\eta -\eta '\gg m^{-1}$, the integral is dominated by the low
frequencies (close to the branch point). In this regime, the
noise kernel becomes

\be
N(\eta ,\eta ')\sim {V\over 4}\sqrt{\pi\over m}\{{\cos [2m(\eta -\eta ')-\pi/4]
\over (\eta -\eta ')^{3/2}}\}
\te

It is tempting to conjecture that the delta function-like singularity in the
noise kernel is due to the contribution of those $\vec x$ such that $(\eta ',
\vec x)$ lies close to the past light cone of the point $(\eta ,t_i)$. Indeed,
for $\eta\ge\eta '$ we have the exact expression

\be
G_+={-m\over 8\pi}(-\sigma^2)^{-1/2}H_1^{(2)}[m(-\sigma^2)^{1/2}]
\te
where $\sigma^2$ is the four dimensional geodesic distance between
the arguments of the propagator. As $\sigma^2\to 0$, this yields a
mass independent leading singularity, which is indeed equivalent to
the whole massless propagator. Thus the integration over $\vec x$ of this term
alone reproduces the delta function-like behavior of the noise kernel.

This result in turn suggests that the delta function-like behavior will be
common to all Hadamard vacua in curved spacetimes, as these share
the singularity structure of the Minkowski propagators
\cite{BirDav}. The details of
the particular evolution are coded in the nonsingular tail of the
noise kernel.

The dissipation kernel for a scalar field in flat space time is analyzed in
Ref. \cite{HuBelgium}, where the fluctuation-dissipation theorem is also
stated.

\newpage
\section{Stochastic Sources in the Energy Density}

The established theory of semiclassical gravity is based on the Einstein
equation for classical metric with a source given by the
the vacuum expectation values of the energy-momentum tensor
of a quantum field.
A major proposal we advance in this paper is that noise terms should
be added as a stochastic source to the semiclassical Einstein equation
beyond the usual order of approximation, turning it into an Einstein-Langevin
equation.
These noise terms arise from the difference between
the average amount of vacuum polarization and particle creation
(measured by the expectation value of the energy momentum tensor) and
the actual value of the same quantities in a specific history.
By comparison to our generalized semiclassical theory
the usual semiclassical Einstein theory can be viewed
as only a mean-field theory.
It is well-known that the mean field theory is inadequate
in addressing the full effect of fluctuations and instability, as in the
studies of critical phenomena \cite{Ma} .
To the extent the transition from classical general relativity
to quantum gravity may involve instability and phase transitions, the old
theory is ill-prepared for such an analysis.
We will discuss the ramifications of this generalized theory
in future works.

In this section we shall explain the nature of noise by relating it to
fluctuations in particle number and vacuum polarization
, using simple models in introductory
quantum field theory in curved spacetimes. We shall calculate
the amount of particle creation in a cosmological evolution
with asymptotically static regions. Even simpler still,
we assume that the evolution never deviates much
from Minkowski space time. We shall do this in two ways, first by deploying
the machinery from the previous sections, and then by a
straightforward analysis based on elementary quantum field theory. We
shall show that both analysis give the same result in their common
range of applicability. This will explain the meaning of the noise term
in the Einstein-Langevin equation.

\subsection{CTP effective action and the energy-momentum tensor}
In the last section we have seen how the variation of the CTP
effective action with respect to the conformal factor yields the trace
of the stress-energy tensor. We also noted that the presence of a stochastic
source in the semiclassical equations for the conformal factor is equivalent to
a random component in the trace of the energy momentum tensor
of the field. In a FRW Universe we also know that the trace determines
the full energy momentum tensor \cite{Par79}.
In particular, the energy density can be related to the trace by

\be
T^0_0(\eta )={1\over a^4}[{\rm constant}+
\int^{\eta}~d\eta '~a^3\dot aT^{\mu}_{\mu}]
\te
Let us now consider an asymptotically-static, near-flat evolution,
where the mode frequencies $\omega^2$ are composed  of  a
constant part $\omega_0^2$, and a small time varying component $\Delta
\omega^2$ (we retain this notation for the fluctuating part for simplicity,
although properly speaking we do not treat it as a perturbation, but as
part of the background).

In the OUT region, where $a=1$ again, the energy density
has a `deterministic' part

\be
\rho_d ={\rm constant}- \int~d\eta d\eta '
{d\Delta\omega^2\over d\eta}(\eta )
D(\eta ,\eta ')\Delta\omega^2(\eta ')
\te
which, after integration by parts, becomes

\be
\rho_d ={\rm constant}+ \int~d\eta d\eta '\Delta\omega^2(\eta )
{\partial D(\eta ,\eta ')\over\partial\eta}\Delta\omega^2(\eta ');
\te
and a `stochastic' part

\be
\rho_s =-\int ~d\eta~ \Delta\omega^2(\eta )
{\partial \xi(\eta )\over\partial\eta}.
\te
Consequently, the mean deviation from the average value is

\be
\langle\rho_s^2\rangle =\int ~d\eta d\eta '~\Delta\omega^2(\eta )
{\partial^2{N}(\eta ,\eta ')\over\partial\eta\partial\eta '}\Delta\omega^2
(\eta  ').
\te
The dissipation kernel $D$ is given by

\be
D(\eta, \eta')=-Im
\{ V\int d^3\vec x~G_+^2((\eta ,0),(\eta ',\vec x))\theta (\eta -\eta ')\}.
\te
Using the representation

\be
\theta (\eta -\eta ')=i\int~{d\omega\over 2\pi}~{e^{-i\omega (\eta -\eta ')}
\over\omega +i\epsilon}
\te
and the formulae from Section {\bf 4.4}, $D$ reduces to

\be
D(\eta ,\eta ')={V\over 64\pi^3}\int~d\omega~e^{-i\omega (\eta -\eta ')}
\int_{4m^2}^{\infty}{dt\over t-(\omega +i\epsilon )^2}\sqrt{1-{4m^2\over t}},
\te
which yields the average energy density

\be
\rho_d={V\over 32\pi^2}\int_{2m}^{\infty}~d\nu~\nu\sqrt{1-{4m^2\over\nu^2}}
|\Delta\omega^2(\nu )|^2
\te
where

\be
\Delta\omega^2(\nu )=\int~d\eta~e^{-i\nu\eta}\Delta\omega^2(\eta ).
\te

In turn, the noise kernel $N(\eta, \eta')$ is

\be
N(\eta, \eta')={1\over 4}
Re \{ V\int d^3\vec x~G_+^2((\eta,0),(\eta ',\vec x))\}
\te
and reduces to

\be
N(\eta, \eta')={V\over (8\pi )^2}\int~d\omega~e^{i\omega (\eta -\eta ')}
\sqrt{1-{4m^2\over\omega^2}}\theta (|\omega | -2m)
\te
which yields the fluctuations in the energy density

\be
\langle\rho_s^2\rangle =
{V\over 32\pi^2}\int_{2m}^{\infty}~d\nu~\nu^2\sqrt{1-{4m^2\over\nu^2}}
|\Delta\omega^2(\nu )|^2.
\te

Since in the asymptotic region there is no vacuum polarization, these
fluctuations can be adscribed solely to fluctuations in the number of created
particles.   We  shall  show  now  that  it  is indeed so, by computing these
fluctuations  independently.      In  this  way,  we  shall  demonstrate  the
consistency of our approach  with  familiar results from quantum field theory
in curved spaces, in this simple example.

\subsection{Fluctuations in Particle Number}

Let us study the same problem, that is, fluctuations in
the energy density in the OUT region of a near-flat, asymptotically-static
evolution, by using simple arguments from
quantum field theory in curved spacetimes.

Physically there is energy in the OUT region because
particles have been created. Indeed they  are created with
a definite spectrum. We know from standard source \cite{cpc}
that the probability of finding $(2n+1)$
particles in mode $k$ vanishes, and the probability of finding $2n$
particles is

\be
p_{2n}={(2n)!\over 2^{2n}(n!)^2}{|\beta |^{2n}\over |\alpha |^{2n+1}}.
\te
With the help of the elementary series

\be
\sum_{n=0}^{\infty}[{(2n)!\over 2^{2n}(n!)^2}]x^{n}={1\over\sqrt{1-x}}
\label{sum}
\te
it is immediate to get the average number of created particles

\be
\langle n\rangle =|\beta |^2
\te
and the fluctuation in number

\be
\langle n^2\rangle -\langle n\rangle^2=2|\alpha |^2|\beta |^2.
\te
(We could also obtain these results by computing the IN vacuum expectation
values of the OUT particle number operator).

Observe that all modes have (positive) frequency greater than $m$,
and that the number of modes with frequencies between $\nu$ and $\nu +d\nu$
is

\be
{4\pi Vk^2dk\over (2\pi )^3}={V\over 2\pi^2}\sqrt{1-{m^2\over\nu^2}}\nu^2
d\nu
\te
so that the average energy density becomes

\be
\rho_d={V\over 2\pi^2}\int_{m}^{\infty}
{}~d\nu\nu^3\sqrt{1-{m^2\over\nu^2}}|\beta |^2
\te
and the fluctuations in energy density are given by

\be
\langle\rho_s^2\rangle ={V\over \pi^2}\int_{m}^{\infty}~d\nu
\nu^4\sqrt{1-{m^2\over\nu^2}}|\alpha |^2|\beta |^2
\te
Although the particle number for a particular mode
is certainly not a Gaussian variable, the energy will be, by virtue of the
central limit theorem.

In the simple case in question, the Bogolubov  coefficients are (see
Sec. 4) approximated by $\alpha \approx 1$, and

\be
\beta (\nu )=({i\over 2\nu })\Delta\omega^2(2\nu )
\te
Introducing this expression in the formulae above, we immediately recover
the results from Subsection 5.1. Of course, we already noted in
Refs \cite{CalHu87,CalHu89} that the energy dissipated from the conformal
factor is equal to the energy of the created particles. The new ingredient
found here is that fluctuations in energy density, which constitutes noise
in the semiclassical Einstein equation,
also have a simple physical interpretation.

The above analysis is only a trivial application of a very powerful tool,
designed to illustrate the meaning of some new aspects in these
methods. Of course,
one doesn't need such heavy formalisms to treat these simple models.
In a more complex problem the new method we are proposing here not
only allows us to compute the fluctuations in the asymptotically-static
regions,
it also tells us how to feed back those fluctuations into the
evolution of the conformal factor, even at intermediate times, where
there may not be a well defined particle number operator. It is also
very important that once the physical
situation is defined (e.g., what is the initial state, and what will be coarse-
grained over), the formalism will generate the correct results with
self-consistency without the need of making {\it ad hoc} assumptions
or adjustments along the way. In some examples given in Sec. 4
we have seen the danger of taking a problem at face-value
in seeking convenient solutions to deeper issues.

\section{Discussion}

Our earlier papers \cite{CalHu87,CalHu89}
showed how the Schwinger-Keldysh (CTP) method can be successfully
applied to treat particle creation and backreaction in semiclassical cosmology.
We obtained a real and causal equation of motion and showed how
particle creation can be viewed as a dissipative process.
In this paper we have developed further this method
to incorporate the treatment of noise and fluctuations, relating them to
decoherence and particle creation.
With the help of the Feynman-Vernon influence functional method we
can understand better the statistical
mechanical meaning of the quantum processes
involved and expound the origin and nature of noise and fluctuations associated
with
quantum fields (and eventually that of spacetime) in semiclassical gravity.
Our findings lead us to propose
a generalized theory of semiclassical gravity where stochastic
source terms corresponding to the fluctuations in the number of
particle creation appear in addition to the usual term corresponding
to the vacuum
expectation value of the energy- momentum tensor.

As is shown here explicitly through some simple examples, these new methods
make it possible to display the full interconnections between particle
creation, decoherence, noise, fluctuation, and dissipation \cite{HuTsukuba}.
Our analysis also confirms previous hints on the connection
between decoherence and particle creation \cite{PazSin,CalMaz},
and the balance
between decoherence and the stability and predictability
of classical evolution \cite{GelHar2}.
Perhaps the most important finding of this work is that semiclassical
evolution is inherently stochastic, and that its statistical
properties may be rigorously derived from the closed-time-path effective
action or the influence/decoherence functional.

{}From a theoretical point of view
these new methods are essential in gaining
a fuller understanding of the intricacies of the relation between quantum and
semiclassical physics, and in particular, semiclassical gravity and cosmology.
Two aspects stand out:
First, their {\it formal structure} bestows upon us a complete
description of the system and its  environment, with almost no room
for any {\it ad hoc} assumptions or piecemeal adjustments. Its logical
extension leads us to new discoveries and insights, such as the noise terms
in the equations of motion or the entropy of the open system from the
reduced density matrix. Second, their {\it intrinsic power}
in the treatment of complex problems like the quantum to classical
transition and cosmological backreaction
which requires a self-consistent
description of particle creation, decoherence, noise, fluctuations
and dissipation on the same footing.

Extending the examples given here to more realistic situations,
this formalism points the way to a more complete and
accurate treatment of problems involving quantum processes
in the very early universe, such as  structure formation and
phase transition problems.
The concept and methodology can also be applied to the study of black hole
thermodynamics in a dynamical setting. The CTP effective action
and the non-equilibrium IF are particularly
suitable to address the backreaction problem of semiclassical black hole
collapse. There, at the verge of the domain of validity of the
semiclassical gravity theory, we expect to see similar stochastic behavior
associated with the quantum field vacuum becoming more prominant as the
gravitational field increases in strength.
As is known from critical phenomena studies, the prominance of the
fluctuation terms signals the onset of instability
(of the ground state) of the old phase-- here described by Einstein's gravity,
and the transiton to the new phase-- possibly described by a theory of
quantum gravity
\cite{Ma}.
We will discuss these issues in future works \cite{HM3,HuSin,CalHuFDR}.

In a broader light, recognition of this unavoidable statistical
feature of semiclasical evolution may affect drastically
our understanding of these `medium energy' phenomena
--which can be viewed in a more general sense as `mesoscopic' physics.
A sketch of these ideas can be found in \cite{HMLA,HuBanff}.
The actual application of these methods to concrete model building
in gravitation and cosmology or other subjects usually hinges upon  the
identification of meaningful coarse grained descriptions adequate to each
particular setting \cite{GelHar2,HuSpain}.
The  results of this paper can help to explicate this issue as well.

On general grounds, the consistent histories approach to quantum physics
assumes
that one has at his/her disposal a fine grained description of the system of
interest, which is subsequently coarse-grained to leave only the physically
relevant variables. However,
what constitutes a meaningful choice of
fine-grained histories depends on the scale of
the problem: a nucleon may be regarded
as an `elementary' particle  at energies of MeVs, while it will reveal its
composite character at energies of GeVs.
Strictly speaking, there is no ground to
believe that an ultimate absolute fine-grained description of the universe
exists, which would seem to void the application of this approach in
cosmological problems.

Our results suggest  a  practical test   to  choose the correct level of
description for a given problem, namely, that
a description of a physical system may
be considered fine-grained insofar as the ever-present dissipative
and stochastic elements in the dynamics are small compared to
the characteristic scale of energy and dimension of observation.
This happens for finer scales of measurement and higher degree of accuracy.
At a coarser scale of measurement or with a lesser degree of accuracy
one effectively averages over certain set of (irrelevant) variables and
would necessarily recognize the appearance of dissipation and fluctuations
phenomena in the coarse-grained description
(see \cite{HuSpain} for further discussions on this point). Thus, for a
given accuracy, it is possible to show rigorously that a given set of
histories can indeed  be treated as fine-grained, even if the underlying
levels of description are unknown.

The application of this criterion along with a generalized concept of
what constitutes a ``history'' (e.g., allowing for collective \cite{HuSpain}
or hydrodynamic \cite{GelHar2} variables, and/or  correlations \cite{CalHuDCH}
as parts of the specification of a history) may help us focus on the key issues
for understanding the nature and origin of the semiclassical regime.

\section*{Acknowledgments}

We thank Andrew Matacz for discussions  and
Juan Pablo Paz and Sukanya Sinha for useful comments
on a preliminary draft of this paper.
EC is partially supported by the Directorate General for Science Research
and Development of the Commission of the European Communities under
Contract N$^o$ C11-0540-M(TT), and by CONICET, UBA and Fundaci\'on
Antorchas (Argentina). BLH is supported in part by the National
Science Foundation (NSF) under grant PHY 91-19726. This collaboration is
partially supported by CONICET and NSF under the Scientific and
Technological Exchange Program between Argentina and U.S.A.

EC wishes to thank D. Hobill and the Kananaskis Center of the University
of Calgary, where parts of this work were completed, for their kind
hospitality. BLH wishes to thank O. Bertolami, M. Bento and J. Mourao
for their warm hospitality during his visit to the Institute Superior Technico
of Lisbon, where part of the results of this work was first reported.

\newpage
\section*{Appendix}

In this appendix, we shall analyze the exact expression (3.12)
for the closed time path effective action given in Sec. 3.
We first give two independent proofs of it (Sec. A.1 and A.2), and then
show that the resulting contribution to the semiclassical equations
of motion is real and causal (Sec. A.3).

\subsection{Proof based on elementary field theory}

This proof is based on the observation that the definition of the
CTP effective action, as given in Section {\bf 2}, may be reduced to

\be
e^{i\Gamma}=\sum_{n}\langle 0IN|nOUT\rangle_-\langle nOUT|0IN\rangle_+
\te
Manipulating the Bogolubov coefficients it is easy to show that

\be
\langle (2n+1)OUT|0IN\rangle =0
\te
and

\be
\langle 2nOUT|0IN\rangle =
[{\sqrt{(2n)!}\over 2^{n}(n!)}]({\beta^{n}\over\alpha^{(2n+1)/2}})^*
\te
(we have chosen the relative phases of the IN and OUT vacua to match
the results from perturbation theory; this choice fixes all the rest).

Using this in the formula for $\Gamma$
and applying the summation formula  Eq. (\ref{sum})
we get the desired result (3.12).

\subsection{Proof based on functional analysis}
\medskip
This proof is based on the functional integral expression

\be
e^{i\Gamma}=\int~D \phi_+D \phi_-~e^{i(S_f[a_+,\phi_+]-S_f[a_-,\phi_-])}
\te
Remember that the positive branch has a negative slope in the complex
$\eta$ plane, and the negative branch has a positive slope. Also that
the CTP histories are continuous across the turning point.

Let $\phi$ be the common value of $\phi_+$ and $\phi_-$ at the
Cauchy surface in the future, and let $\bar \phi_{\pm}$ be the classical
solution, in each branch, that vanishes in the past and matches $\bar \phi$
in that Cauchy surface (we are assuming we have already decomposed
the integral by Fourier modes, so we have a 0 dimensional field theory).
Then each history can be written as $\phi_{\pm}=\bar \phi_{\pm}+\varphi_{\pm}$,
where $\varphi_{\pm}$ vanishes both in the past and the future, and we get

\be
e^{i\Gamma}=\int~d \phi~e^{i(S_f[a_+,\bar \phi_+]-S_f[a_-,\bar \phi_-])}
\int~D\varphi_+~e^{i(S_f[a_+,\varphi_+])}
\int~D\varphi_-~e^{-i(S_f[a_-,\varphi_-])}
\te
Because it is quadratic, the action on a classical solution reduces to

\be
S_f[a_{\pm},{\bar\phi}_{\pm}]=\ha [{\bar\phi}_{\pm} \dot {\bar\phi}_{\pm}
(+\infty)-{\bar\phi}_{\pm} \dot {\bar\phi}_{\pm}(-\infty)]
\te
But because of the boundary conditions, this yields

\be
S_f[a_+, {\bar\phi}_+]-S_f[a_-,{\bar\phi}_-]=\ha \phi
[\dot {\bar\phi}_+-\dot{\bar\phi}_-]
\te

Let $h_{\pm}$ be solutions of the Klein- Gordon equation on each
branch, vanishing in the past (i. e., $h_+$ is  negative frequency,
and $h_-$ is positive frequency) and satisfying an arbitrary normalization
(e. g., having unit Klein-Gordon norm). If the turning point is
located at $\eta =\eta^o$,

\be
\bar \phi_{\pm}=\phi~({h_{\pm}(\eta )\over h_{\pm}(\eta^o)}).
\te
The integral over $\phi$ in the expression for $\Gamma$ is then an
ordinary Gaussian integral, which yields

\be
[{1\over 2\pi i}({\dot h_+(\eta^o)\over h_+(\eta^o)}-{\dot h_-(\eta^o)\over h_-
(\eta^o)})]^{-1/2}.
\te
The remaining functional integrals are of the usual IN-OUT type. Following
Ref. \cite{Gel'fand}, we know that, e.g.,

\be
\int~D\varphi_{\pm}~e^{i(S_f[a_{\pm},\varphi_{\pm}])}=
[Det {\d^2S_f\over
\d \phi_{\pm}^2}]^{-1/2}
\te
And that, in turn

\be
[Det {\d^2S_f\over
\d \phi_{\pm}^2}]={\rm constant}~h_{\pm}(\eta^o)
\te

Using these expressions for the remaining functional integrals, we are led
to

\be
\Gamma =({i\over 2})\ln [h_-(\eta^o)\dot h_+(\eta^o)-h_+(\eta^o)\dot h_-
(\eta^o)]+{\rm constant}
\te
Finally, rewriting the IN functions $h_{\pm}$ in terms of the OUT
particle model and the Bogolubov coefficients, we get the desired formula.

\subsection{Proof of the causal and real property}

We now present the argument why the equations of motion
deriving from the effective action
are necessarily real and causal. The contribution from $\Gamma$
to the equations of motion takes the form

\be
(i/2) [\alpha {\partial\alpha^*\over\partial a(\eta )}-
\beta {\partial\beta^*\over\partial a(\eta )}]
\te
Reality follows from the identity $|\alpha |^2-|\beta |^2=1$, which implies

\be
 Re[\alpha {\partial\alpha^*\over\partial a(\eta )}-
\beta {\partial\beta^*\over\partial a(\eta )}]=0
\te
Thus, the equations of motion are $i$ times something purely imaginary,
and therefore something real.

To see that the equations are causal, we need to show that they are not
changed if we perturb the evolution at the future of time $\eta$. To see
this, assume we choose a particle model at some time $\eta '\ge\eta$, but
still earlier than the perturbation. Then the IN-OUT Bogolubov
coefficients can be expressed in terms of the Bogolubov coefficients
between the IN particle model and the $\eta '$ model, and those linking
the $\eta '$ model to the OUT one:

\be
\alpha =\alpha_{IN,\eta '}\alpha_{\eta ',OUT}
+\beta_{IN,\eta '}\beta_{\eta ',OUT}^*
\te

\be
\beta =\alpha_{IN,\eta '}\beta_{\eta ',OUT}
+\beta_{IN,\eta '}\alpha_{\eta ',OUT}^*
\te
Using this in the equations of motion, observing the $\eta '$-OUT
coefficients do not depend on the metric before $\eta '$, we get

\be
(i/2) [\alpha_{IN,\eta '} {\partial\alpha_{IN,\eta '}^*
\over\partial a(\eta )}-
\beta_{IN,\eta '} {\partial\beta_{IN,\eta '}^*\over\partial a(\eta )}]
\te
which in turn does not depend on the metric after $\eta '$. Since $\eta '$
may be chosen arbitrarily close to $\eta$, this proves causality.
\newpage


\end{document}